\documentclass[aps,twocolumn]{revtex4}

\usepackage{graphicx}
\usepackage{dcolumn}
\usepackage{bm}

\begin{document}

\title{Gauge invariant fermion propagator in
QED$_3$}

\author{M. Franz}
\affiliation{Department of Physics and Astronomy,
University of British Columbia, Vancouver, BC, Canada V6T 1Z1}
\author{Z. Te\v{s}anovi\'c and O. Vafek}
\affiliation{Department of Physics and Astronomy,
Johns Hopkins University, Baltimore, MD 21218}
\maketitle

In a recent preprint Khveshchenko \cite{k1} questioned the validity of
our computation of gauge invariant fermion propagator in
QED$_3$, which we employed as an effective theory of 
high-$T_c$ cuprate superconductors \cite{ftv}. 
We take this opportunity to further clarify
our procedure and to show that the criticism voiced in Ref.\ \cite{k1} is 
unwarranted.

First, we elaborate on the gauge invariance of Brown's propagator
\cite{brown1} and then we argue why this object should be associated 
with the gauge invariant electron 
propagator in our QED$_3$ theory \cite{ftv}. Ordinary fermion
propagator in QED$_3$, $G(x-x')$, is a gauge variant quantity and as such it
is only meaningful when calculated in a particular gauge. Within the 
family of covariant gauges, parameterized by the gauge fixing parameter $\xi$,
we may write this quantity as
\begin{equation}
G_\xi(x-x')=e^{-F_\xi(x-x')}\left[{\left\langle G[x,x';a]\right\rangle_\xi
\over \left\langle \exp{(i\int_x^{x'}a\cdot dr)}\right\rangle_\xi}\right],
\label{c1}
\end{equation}
where $e^{-F_\xi(x-x')}=\langle\exp{(i\int_x^{x'}a\cdot dr)}\rangle_\xi$
and $G[x,x';a]$ is a fermion propagator for given fixed configuration of 
field $a$ defined in Appendix C of \cite{ftv}. 
Subscript $\xi$ reminds us that averages over the gauge field are to be 
taken in the covariant gauge parameterized by $\xi$. Eq.\ (\ref{c1}) is an
identity, but a very useful one. We observe that the expression in the angular
brackets, denoted by $\tilde{\cal G}(x-x')$ in Ref.\ \cite{ftv}, is 
independent of $\xi$ by virtue of the fact that numerator and
denominator transform in the same way, as long as {\em the same} gauge 
transformation is applied to both. Therefore the gauge
dependence cancels out, leaving behind a {\em fully} gauge
independent quantity. This is confirmed by an explicit 
computation of $G_\xi$ and $F_\xi$  which shows that the product
$G_\xi(x-x') e^{F_\xi(x-x')}$ is $\xi$-independent \cite{ftv}. 

In the above sense $\tilde{\cal G}(x-x')$ is gauge invariant. We may now
ask what is the physical content of this propagator? Intuitively,
we might suspect that $\tilde{\cal G}$ is related to
\begin{equation}
{\cal G}(x-x') = \left\langle G[x,x';a]\exp{(-i\int_x^{x'}a\cdot dr)}
\right\rangle ~,
\label{c11}
\end{equation}
which is the natural candidate for the 
electron propagator within our theory. To elucidate this relationship
we exploit the fact that the {\em manifest} gauge invariance of
the Brown's propagator $\tilde{\cal G}(x-x')$ extends {\em beyond}
covariant gauges to the more general family of {\em linear} gauges
specified by the condition ${\cal O}_\mu a_\mu (x) - w(x)=0$,
where three-vector ${\cal O}_\mu$ is a linear differential operator and 
$w(x)$ is an arbitrary scalar function \cite{zj,peshkin}. Observe that
both Landau (${\cal O}_\mu = \partial_\mu$, $w =0 $) and 
axial (${\cal O}_\mu=n_\mu$, $w=0$) gauges 
belong to this more general family. For the purposes
of this note we consider
\begin{equation}
\tilde{\cal G}(x-x')={\left\langle G[x,x';a]\right\rangle
\over \left\langle \exp{(i\int_x^{x'}a\cdot dr)}\right\rangle}~~,
\label{c2}
\end{equation}
where the averages in numerator and denominator are now evaluated 
without restriction on the gauge field -- this is just the
Brown's propagator in the unitary gauge. We 
perform the Fadeev-Popov procedure on the numerator and 
denominator by introducing unity disguised as 
functional integral \cite{peshkin}
\begin{equation}
1= \int {\cal D} \alpha \delta (\Omega (a^\alpha))\det
\left(\frac{\delta\Omega (a^\alpha)}{\delta\alpha}\right)
\label{c2'}
\end{equation}
in both. $\Omega (a^\alpha) = {\cal O}_\mu a^\alpha_\mu (x)-w(x)$ 
specifies the linear
gauge and $ a^\alpha_\mu = a_\mu + \partial_\mu\alpha$.

The functional determinant 
$\det\left(\delta\Omega (a^\alpha) / \delta\alpha\right)$
is {\em independent} of $a_\mu$ for linear gauges and
can be canceled out as long as numerator and
denominator are computed within {\em the same} gauge,
set by $\Omega$ (\ref{c2'}). Furthermore, 
the change of variables $a_\mu\to a_\mu - \partial_\mu\alpha$
eliminates the $\alpha$-dependence 
from the $\delta$-function in (\ref{c2'})
leading to decoupling of the integral over $\alpha(x)$
from the average over the gauge field. 
This decoupled integral,  $\int{\cal D}\alpha e^{i[\alpha(x')-\alpha(x)]}$,
is {\em identical} for both numerator and denominator and cancels
out. Finally, we obtain
\begin{equation}
\tilde{\cal G}(x-x')={\left\langle G[x,x';a]\right\rangle_\Omega
\over \left\langle \exp{(i\int_x^{x'}a\cdot dr)}\right\rangle_\Omega}~~,
\label{c2''}
\end{equation}
where subscript $\Omega$ indicates that the average
over the gauge field is to be evaluated 
subject to an arbitrary linear gauge condition
enforced by $\delta (\Omega (a))$ \cite{remark2}. In Landau gauge, 
$\Omega (a)=\partial_\mu a_\mu$, Eq. (\ref{c2''}) 
tells us something we already know: $\tilde{\cal G}(x-x')$
is the same in unitary and Landau gauges, both being
covariant gauges. In axial gauge, however, we have
$\Omega (a)=n_\mu a_\mu$, where $n_\mu$ is a constant three-vector,
and we are free to choose  $n_\mu$ along the direction of the
line integral in (\ref{c2''}). The line integral then vanishes
and we recognize the right hand side of (\ref{c2''}) in
this axial gauge as being equal to ${\cal G}(x-x')$
(see also Appendix C of \cite{ftv}). Since  
$\tilde{\cal G}(x-x')$
and ${\cal G}(x-x')$ are both gauge invariant objects, we
must conclude they are equal in all gauges, linear or not. 
This result is consistent with the argument of Ref.\ \cite{brown1}.

The above procedure, employed in Appendix C of \cite{ftv},
appears reasonably straightforward if perhaps somewhat formal \cite{remark1}. 
Consider here yet another, more physical argument.
In principle, we could try to evaluate the electron propagator
(\ref{c11}) directly. The problem,
of course, is that ${\cal G}$ is ill-defined due to the
UV divergence of the line 
integral and all attempts at direct computation
thus far have yielded a rather problematic
{\em negative} anomalous dimension. Imagine,
however, that we knew the {\em true} form of
${\cal G}(x-x') = \langle G[x,x';a]\exp(-i\theta[x,x';a])\rangle$,
where $\theta[x,x';a]$ is some unknown
phase factor functional which transforms correctly
under gauge transformations, whose UV behavior 
is regular and whose long
distance properties are faithfully represented by the line integral.
Then, in the spirit of (\ref{c1}), 
we can define the generalized Brown's propagator 
$\tilde{\cal G}^\theta (x-x')$:
\begin{equation}
G_\xi(x-x')=\langle \exp(i\theta[x,x';a])\rangle_\xi 
\tilde{\cal G}^\theta (x-x')~~.
\label{c3}
\end{equation}
In the generalized Yennie gauge, defined 
as $\langle \exp(i\theta[x,x';a])\rangle_\xi = 1$, \cite{lro} we
have $ \tilde{\cal G}^\theta = G_\xi$.
On the other hand, the true electron propagator
in the generalized Yennie gauge also 
equals:
\begin{eqnarray}
{\cal G}(x-x') = \langle G[x,x';a]\exp(-i\theta[x,x';a])\rangle\to 
\nonumber \\
\to \langle G[x,x';a]\rangle_\xi \langle\exp(-i\theta[x,x';a])\rangle_\xi
= G_\xi(x-x') ~~.
\label{c4}
\end{eqnarray}
Although in general the above decoupling is not valid,
we can perform it in the generalized Yennie gauge 
since $\langle\exp{(i\theta ) }\rangle_\xi =1$
implies that the phase factor has no fluctuations
(or, alternatively, has ``long range order'' \cite{lro})
and can be pulled outside the average $\langle\cdots\rangle$.
Consequently, the Brown's propagator $\tilde{\cal G}^\theta (x-x')$ and the
true electron propagator ${\cal G}(x-x')$ coincide in the
generalized Yennie gauge. 

The above ``physical'' argument falls short of the precise
mathematical theorem. For example, one can question
whether the generalized Yennie gauge actually exists,
or whether $\exp(i\theta[x,x';a])$ itself exists,
and so on. Such objections, however,
are as impractical to dispel as they are pointless.
Within the dimensional regularization method adopted
in our paper, $\exp(i\theta[x,x';a])$ obviously
exists, it {\em is}
represented by the line integral, and the generalized Yennie
gauge {\em does} exist, at $\xi =2$. The internal
logic of our method then demands we identify
$\tilde{\cal G}$ with ${\cal G}$, just as we did. 

To summarize, by following a standard procedure \cite{peshkin},
we have computed the electron propagator in QED$_3$ theory \cite{ftv}.
The form of this propagator appears reasonable on physical grounds.
This result of Ref. \cite{ftv} stands as is.


\begin{thebibliography}{10}
\bibitem{remark2} At this point we could perform the usual functional 
integration over $w(x)$ with a Gaussian measure parametrized by gauge fixing
parameter $\xi$ \cite{peshkin} and demonstrate that $\tilde{\cal G}$ remains
the same in all such gauges. To keep discussion simple we confine ourselves
to Landau and axial gauges which is all we need to prove our point.
\bibitem{lro} More generally, we seek a gauge in which 
$\exp(-i\theta[x,x';a])$ has ``long range order'', i.e.
$\langle \exp(-i\theta)\rangle\to {\rm const.} < 1$ for
$|x-x'|\to\infty$. One can then show that 
$\tilde{\cal G}$ and ${\cal G}$ share the same leading 
power law behavior and are therefore ``equal'' for
the purposes of our paper.
\end{thebibliography}
\end{document}